
\hoffset0cm
\voffset-1.5cm
\tolerance=2000
\hbadness=2000
\vbadness=10000
\parindent=20.pt

\font\bigfett=cmbx10 scaled\magstep2

\def\pslash{\not{\hbox{\kern-2.3pt $p$}}}
\def\qslash{\not{\hbox{\kern-2.0pt $q$}}}
\def\lslash{\not{\hbox{\kern-0.5pt $l$}}}
\def\rslash{\not{\hbox{\kern-2.3pt $r$}}}

\magnification=\magstep1
\hsize 13cm
\baselineskip=16pt plus .5pt minus .5pt

\rightline{MZ-TH/92-25}
\rightline{TTP92-23}
\rightline{May 1992}
\vskip 0.8cm
\centerline {\bigfett DOMINANT THREE-BODY DECAYS}\bigskip
\centerline {\bigfett OF A HEAVY HIGGS AND TOP QUARK}\bigskip
 \bigskip\bigskip\bigskip
\centerline {\bf Roger~Decker, Marek~Nowakowski}
\centerline {Institut f\"ur Theoretische Teilchenphysik }
\centerline {Universit\"at Karlsruhe }
\centerline {Postfach 6980 }
\centerline {D-7500 Karlsruhe 1, Germany} \bigskip
\centerline{$and$}\bigskip
\centerline {\bf Apostolos~Pilaftsis}
\centerline {Institut f\"ur Physik }
\centerline {Johannes-Gutenberg Universit\"at}
\centerline {Postfach 3980 }
\centerline {D-6500 Mainz, Germany} \bigskip
\centerline{ to appear in Z.~Phys.~C}
\bigskip
\centerline{\bf ABSTRACT}\smallskip
We calculate the dominant three body Higgs decays, $H \to W^+W^-(Z^0,
\gamma)$ and $H \to t\overline{t}(Z^0,\gamma ,g)$, in the Standard Model.
We find that the branching ratios of these decays are of the order of few
percent for large Higgs masses. We comment on the behaviour of the partial
decay width $\Gamma (H \to t\overline{b}W^-)$ below the $t\overline{t}$
threshold. Numerical results of the following three body top decays, $t \to
W^+b(\gamma ,g,Z^0)$ and $t \to W^+bH$, are also given. We discuss the
feasibility of observing these Higgs and top decays at future high energy
colliders.
\vfill\eject
{\bf 1. Introduction} \medskip
In spite of the beautiful confirmation of the Standard Model (SM) by
LEP data [1] the Higgs sector [2] still remains quite unconstrained. The only
`real' upper bound on the Higgs mass is $M_H \le 1TeV$ given to us by
unitarity arguments [3]. On the other hand if the Higgs particle is indeed
heavy, $M_H \ge
500GeV$, then three body Higgs decays like $H \to W^+W^-(\gamma ,Z^0)$ and
$H \to t\overline{t}(Z^0, \gamma ,g)$ have appreciable branching ratios
too.
In this case in order to confirm that an observed scalar (should one find
it) is indeed the Higgs particle predicted by SM one will have to
study
the subdominant partial decays such as those mentioned above, as well.
\smallskip
As is well known two body decays of a heavy Higgs decays are dominated by $H
\to
W^+W^-, Z^0Z^0, t\overline{t}$ . Some of the more
important three body decays have already been discussed in the literature.
The pure bosonic decay modes $H \to W^+W^-Z^0$ and $H \to W^+W^- \gamma$
have been calculated in refs. [4] and [5], [6], respectively. Ref. [6]
corrects the results of [5]. We have redone the calculation for both these
channels. For the latter we agree with [6], however we disagree with ref.
[4] for the former by two orders of magnitude. Furthermore a heavy top
means a substantial Yukawa coupling $g_{t\overline{t}H}$. Due to this fact
the decays $H \to t\overline{t}(\gamma ,g,Z^0)$ are also not negligible.
Only one of them ($H \to t\overline{t}Z^0$) has been discussed so far
[7].
Here we take the opportunity to correct an omission of factor $3$ in
[7]
and present the full results for all the three decays.
\smallskip
By now it is clear that the top quark is likely to be heavy ($90\ GeV
\le m_t \le 200\ GeV)$). Therefore its three body decays become
interesting, too. Here we calculate $t \to W^+b(\gamma ,g)$, $t \to
W^+bZ^0$ and $t \to W^+bH$. The first two  have been
discussed by two theoretical groups, which reported similar results [8,9].
Our numerical results agree with ref.~[9] and the revised results of~[8].
\smallskip
In view of the discrepancies that existed between different calculations
for the three body decays of $H$ and $t$ we take the opportunity to
consolidate and to present some new results for all these decays.
\smallskip
Our paper is organized as follows. In section 2 we discuss three body
Higgs decays. Section 3 deals with the corresponding top decays.
Whenever the result for the matrix element squared is not too unwieldy
we present the analytical expression as well. We summarize our conclusions
in the end.
\vfill\eject
{\bf 2. Dominant three body Higgs decays} \medskip
We consider here the following partial decays of a heavy Higgs
$$\eqalign{&H \to W^+W^-(\gamma ,Z^0) \cr
&H \to t \overline{t}(Z^0 ,g, \gamma) \cr
&H \to t\overline{b} W^- \cr}\eqno(1)$$
The corresponding diagrams are depicted in figs. 1 and 2, respectively.
All these decays listed above gain from the fact that the Higgs couples
predominantly to heavy particles e.g.
$$\eqalign{&g_{{}_{HWW}}=gM_W    \cr
&g_{{}_{HZZ}}=g_{{}_{HWW}}\displaystyle{M_Z^2 \over M_W^2}   \cr
&g_{{}_{Ht\bar t}}=g \displaystyle{m_t \over 2M_W} \cr}\eqno(2)$$
The interesting feature of the first decay channel in (1) are the
triple gauge boson couplings.
The trilinear vertices for $W^+(p)-W^-(q)-V^0(r)$ ($V^0=\gamma,Z^0$) are given
by
$$\eqalign{V_{\rho \mu \nu}^{WWV}(p,q,r)&=g_{{}_{VWW}}\left[(r-q)_{\rho}
g_{\mu \nu}\ +\ (q-p)_{\nu}g_{\rho \mu}\ +\ (p-r)_{\mu}g_{\nu \rho}\right]
\cr
&g_{{}_{\gamma WW}}=e,\   g_{{}_{ZWW}}=g\cos\theta_{W} \cr}\eqno(3)$$
which have been defined for all momenta incoming. Additionally, non-standard
couplings like the anomalous magnetic moment vertex
$$ie(\kappa_V -1)\left[g_{\rho \nu}r_{\mu}-g_{\mu \nu}r_{\rho}\right]\eqno(4)$$
and others [10] will be not taken into account here. We think that
a rare decay mode of Higgs is not the best place to look for such
couplings.
\smallskip
Since we consider here a heavy Higgs we have to a very good
approximation
$$\eqalign{&\Gamma_{tot}^{Higgs}=\Gamma(H \to W^+W^-)+ \Gamma(H \to Z^0Z^0)
+ \Gamma(H \to t\overline{t})\cr
&\Gamma(H \to VV)={\alpha_w \over 16n_V}M_Hx_V^{-2}(1-4x_V^2)^{1/2}(1-
4x_V^2+12x_V^4),\ \ \ V=Z^0,W\cr
&\Gamma(H \to t\overline{t})={3 \alpha_w \over 8} M_H\left({m_t \over M_W}
\right)^2(1-4x_t^2)^{3/2} \cr}\eqno(5)$$
where $\alpha_w=g^2/4\pi$, $x_i=M_i/M_H$ and $n_V$ equals $1$($2$) for the
$W^+W^-$ ($Z^0Z^0$) boson pairs. \smallskip
The partial width of the three body Higgs decay can generally be
put into the form
$$\Gamma(H \to 3body)=\displaystyle{1 \over 256\pi^3
M_H^3}\int_{s_2^-}^{s_2^+}ds_2
\int_{s_1^-}^{s_1^+}ds_1\overline{|T(H \to 3body)|^2}\eqno(6)$$
where $s_1$ and $s_2$ are invariants (and $s_{1,2}^{\pm}$ the
corresponding phase space boundaries) which we define below  for each case
under consideration.
\smallskip
Throughout the paper we will use the following set of parameters
$$\eqalign{&\alpha_{em}(M_W^2)=\displaystyle{1 \over 128} \cr
&\alpha_s(M_Z^2)=0.12 \cr
&M_W=80.6\ GeV \cr
&M_Z=91.161\ GeV \cr}\eqno(7)$$
\bigskip
{\bf 2.1 $H \to W^+W^-(Z^0,\gamma)$}
\medskip
The sum of the three diagrams for the decay $H(p) \to W^+(k_+)W^-(k_-)Z^0(k
)$ (see fig. 1) can be conveniently written as
$$T(H \to
W^+W^-Z^0)=-ig_{{}_{ZWW}}g_{{}_{HWW}}\left(T_A^{WWZ}+T_B^{WWZ}+T_C^{WWZ}
\right)\eqno(8a)$$
$$\eqalign{&T_A^{WWZ}+T_B^{WWZ}+T_C^{WWZ}=\displaystyle{\epsilon_{\rho}^*(k_-)
\epsilon_{\mu}^*(k) \epsilon_{\alpha}^*(k_+) \over \Delta_W(s_1) \Delta_W(s_2)
\Delta_Z(s_3)}\ast \cr
&\bigg\{-k^{\rho}g^{\alpha \mu}\Delta_Z(s_3)\left[2\left(\Delta_W(s_1)
+\Delta_W(s_2)\right)-\xi_Z \Delta_W(s_2)\right] \cr
&+k_-^{\mu}g^{\alpha \rho}\Delta_W(s_1)\left[2\Delta_Z(s_3)+\xi_Z \Delta_W
(s_2) \right] \cr
&+k^{\alpha}g^{\mu \rho}\Delta_Z(s_3) \left[2\left(\Delta_W(s_1)+\Delta_W(
s_2)\right)-\xi_Z \Delta_W(s_1)\right] \cr
&-k_+^{\mu}g^{\alpha \rho}\Delta_W(s_2)\left[2\Delta_Z(s_3)+\xi_Z \Delta_W
(s_1)\right] \cr
&-k_-^{\alpha}g^{\mu \rho} \xi_Z\Delta_W(s_1)\left[\Delta_Z(s_3)+\Delta_W
(s_2)\right] \cr
&+k_+^{\rho}g^{\alpha \mu} \xi_Z \Delta_W(s_2)\left[\Delta_Z(s_3)+
\Delta_W(s_1)\right]\bigg \}\cr}\eqno(8b)$$
where we have used eq.(2) and defined
$$\eqalign{&\Delta_i(s)=s-M_i^2 \cr
&\xi_Z=\displaystyle{M_Z^2 \over M_W^2} \cr
&s_1=(k_+ + k)^2 \cr
&s_2=(k_- + k)^2 \cr
&s_3=M_H^2+M_Z^2 +2M_W^2-s_1 -s_2 \cr}\eqno(9)$$
The phase space boundaries which correspond to these variables
are as follows
$$\eqalign{s_2^+&=(M_H-M_W)^2, \ \ \ s_2^-=(M_W+M_Z)^2 \cr
s_1^{\pm}&=M_W^2+M_Z^2-\displaystyle{1 \over 2s_2} \bigg[(s_2-M_H^2+M_W^2)(
s_2+M_Z^2-M_W^2)\cr
&\mp \lambda^{1/2}(s_2,M_H^2, M_W^2)\lambda^{1/2}(s_2,
M_W^2, M_Z^2)\bigg] \cr
\lambda(x,y,z)&=(x-y-z)^2-4xy \cr}\eqno(10)$$
It seems unreasonable to give here the lengthy expression for $|T(H \to W^+W^-
Z^0)|^2$ in full detail. We used the algebraic manipulation program
REDUCE to evaluate this expression and performed the integration
numerically. The results are presented in fig.3. One can see that at
about $M_H \ge 600\ GeV$ the branching ratio is
$$Br(H \to W^+W^- Z^0) \simeq {\cal O}(1\%)\eqno(11)$$
which is
indeed sizeable.
\smallskip
We will discuss the visibility of each of the three body Higgs decays
at the end of section 2.
\smallskip
Our result disagrees with ref.[4] where only the mode $H \to W_LW_LZ_L$
has been evaluated with claim that this is the biggest contribution. The
estimate given in [4] is
$$\displaystyle{\Gamma(H \to W_LW_LZ_L) \over \Gamma(H \to WW)+\Gamma(H \to
ZZ)}=3.07\ast10^{-4}(M_H^2/1TeV)\eqno(12)$$
Our results e.g. for $M_H=1TeV$ is about 2 orders of magnitude larger
which indicates that $W_LW_LZ_L$ is not yet the dominant contribution
to $H \to WWZ$ (in contrast what one would naively expect)
\footnote{$^1$}{After our manuscript has been completed we become aware
of ref. [11] where it is nicely explained why the decay $H\to
W_LW_LZ_L$ is forbidden due to parity conservation on the tree level.
Our results agree with the ones obtained in [11].}.
\smallskip
In case of $H(p) \to W^+(k_+)W^-(k_-)\gamma(k)$ we have only two amplitudes
$T_A^{WW
\gamma}$ and $T_B^{WW\gamma}$ (fig.1). This simplifies the form of the
matrix element considerably and it is given by
$$\eqalign{T(H \to W^+W^-\gamma)=-iegM_W\epsilon_+^{\mu *}\epsilon_-^{\rho *}
\epsilon^{\nu *}\bigg\{&\displaystyle{p \cdot k \over
(k_+ \cdot k)(k_- \cdot k)} \left(k_{\rho}g_{\mu \nu} -k_{\mu}g_{\rho \nu}
\right)\cr
& +g_{\mu \rho}\left[\displaystyle{k_{+ \nu} \over k_+ \cdot k}
-\displaystyle{k_{- \nu} \over k_- \cdot k}\right]\bigg\}\cr}\eqno(13)$$
Written in this way, the amplitude is trivially transverse ($k_\nu
T^{\mu\rho\nu}=0$.
This simple form enables us to express $|T(H \to W^+W^-\gamma)|^2$ in a
relatively compact form given in the appendix (eq.(A2)). The variables
$s_1$ and $s_2$ as well as the phase space limits are obtained from
eqs.(10)
by putting formally $M_Z=0$. To avoid infrared singularities we
impose a cut on the photon energy which in terms of our integration
variables is
$$E^{\gamma}=\displaystyle{k \cdot p \over M_H}=\displaystyle{s_1+s_2
-2M_W^2 \over 2M_H} > E^{\gamma}_{cut}\eqno(14)$$
In the numerical calculation we have chosen $E^{\gamma}_{cut}=10,20,50
\ GeV$
corresponding to the full, dashed and dotted lines in fig.4, respectively.
As in the case $H \to W^+W^-Z^0$ the branching ratio $Br(H \to W^+W^-\gamma
)$ is also of the order ${\cal O}(1\%)$ and hence not negligible. Our
calculation confirms the results of ref.[6] which in turn corrected the
mistake in [5]. We also mention that a detailed treatment of $H\to
WW\gamma$ for the case of soft
bremsstrahlung (as a part of radiative corrections) as well as for
hard photon emission is given in ref. [12]. There a partial
analytical integration over the phase space (in the range $E^\gamma_{\rm cut}
<< E^\gamma <E^\gamma_{\rm max}$ can also be found. We have checked
that our numerical results agree with fig. 6 of ref. [12].
\bigskip
{\bf 2.2 $H \to t\overline{t}(Z^0, \gamma ,g)$}
\medskip
In this section we will mainly concentrate on the decay channels
$H \to t\overline{t}(\gamma ,g)$. The partial decay $H \to t\overline{t}Z^0$
has been calculated by us in [7]. Note the in [7] an overall colour
factor $3$ is missing which we have now included in the numerical results
for this channel presented in fig.5. For $M_H \ge 600\ GeV$ the branching
ratio for this particular decay mode is of the order $10^{-3}$ and higher.
\smallskip
It is also evident that once the process $H \to t\overline{t}\gamma$ has
been calculated the corresponding gluonic mode $H \to t\overline{t}g$
can be obtained from the former by including the Casimir factor of
$SU(3)_C$ (and reducing it by $1/N_C$)
$$\displaystyle{1 \over N_C} {\rm Tr}\left(\displaystyle{\lambda^a \over 2}
\displaystyle{\lambda^a \over 2}\right) =\displaystyle{4 \over 3}\eqno(
15)$$
and by replacing
$$\alpha_{em} \to \displaystyle{9 \over 4}\alpha_s \eqno(16)$$
This leads to
$$\eqalign{\Gamma(H \to t\overline{t}g)&=3\displaystyle{\alpha_s(M_H^2)
\over \alpha_{em}}\Gamma(H \to t\overline{t}\gamma) \cr
&\simeq 38.4 \Gamma(H \to t\overline{t}\gamma) \cr}\eqno(17)$$
The total matrix element $T(H \to t\overline{t}\gamma)$ is given by
the sum of the two amplitudes
$$\eqalign{&T_A^{t\overline{t}\gamma}=-\displaystyle{i \over 2}
\epsilon^{\mu *}(k_{\gamma})
\displaystyle{ge \over s_1 -m_t^2}\displaystyle{m_t \over M_W}
\overline{u}_t\left(\lslash+m_t \right)\gamma_{\mu} v_{\overline{t}} \cr
&T_B^{t\overline{t}\gamma}=-\displaystyle{i \over 2}
\epsilon^{\mu *}(k_{\gamma})
\displaystyle{ge \over s_3 -m_t^2}\displaystyle{m_t \over M_W}
\overline{u}_t \gamma_{\mu}\left(\rslash +m_t \right)v_{\overline{t}} \cr}
\eqno(18)$$
where
$$\eqalign{&l=-k_{\gamma}-k_{\overline{t}},\ \ \ l^2=s_1 \cr
&r=k_{\gamma}+k_t,\ \ \ \ \ r^2=s_3 \cr
&s_3=M_H^2+2m_t^2-s_1-s_2 \cr}\eqno(19)$$
The phase space boundaries to be used in the numerical evaluation
of the eq.(6) are given by
$$\eqalign{&s_2^+=M_H^2,\ \ \ s_2^-=4m_t^2 \cr
&s_1^{\pm}=\displaystyle{1 \over
2}\left(M_H^2+2m_t^2-s_2\right)\pm\displaystyle
{1 \over 2}\left(M_H^2-s_2 \right)\left[1-4\displaystyle{m_t^2 \over s_2}
\right]^{1/2} \cr}\eqno(20)$$
After factorizing common constants we put the squared matrix element
into the following form
$$\overline{|T(H \to t \overline{t}\gamma)|^2}=\displaystyle{64 \over 3}
\pi^2 \alpha_w \alpha_{em}\displaystyle{m_t^2 \over M_W^2}\left(|\tilde{T}
_A^{t\overline{t}\gamma}|^2+|\tilde{T}_B^{t\overline{t}\gamma}|^2
+2{\it Re}\tilde{T}_A^{t\overline{t}\gamma} \cdot
(\tilde{T}_B^{t\overline{t}\gamma})^*
\right)\eqno(21)$$
and refer the reader to the appendix where the analytical form is given
(eq.(A3)).
\smallskip
Fig.6a and 6b show the numerical results for two different
values of $E^{\gamma}_{cut}$ imposed on the photon energy. The branching
ratio is of the order $10^{-3}$. From this using eq.(17) we conclude that
$$Br(H \to t\overline{t}g)\simeq {\cal O}(1\%)\eqno(22)$$
which is comparable to $Br(H \to W^+W^-Z^0)$.
\smallskip
Analytical expressions in term of Spence functions for hard
bremsstrahlung in $ H\to f\overline{f} \gamma$ and
$H\to q\overline{q} g$ are presented in [13] and [14], respectively.
\bigskip
{\bf 2.3 $H \to t \overline{b} W^-$}
\medskip
This decay relevant for the kinematical range $M_H \le 2m_t$ has been
discussed in detail in [7]. We supplement here this discussion by giving
in the appendix (eqs.(A4)-(A6)) the analytical expression for
$|T(H \to W^-t\overline{b})|^2$ which we write in the form
$$\overline{|T(H \to W^-t\overline{b})|^2}=6 \alpha_w \pi^2 |V_{tb}|^2
\left(|\tilde{T}_A^{Wtb}|^2 +
|\tilde{T}_C^{Wtb}|^2 +2{\it Re}(\tilde{T}_A^{Wtb} \cdot (\tilde{T}_C^{Wtb}
)^* \right)\eqno(23)$$
Note that a third amplitude in which the Higgs couples directly to
bottom-quarks can be safely neglected since
$$|T_B(H \to \overline{b}b^* \to \overline{b}W^+t)|^2 \simeq \left({m_b \over
m_t}\right)^2|T_A^{Wtb}|^2 < 2.10^{-3}|T_A^{Wtb}|^2 \eqno(24)$$
The tilded quantities in eq.(23) depend on the two Mandelstam variables
$$\eqalign{&s_1=(k_W+k_b)^2 \cr
&s_2=(k_t+k_b)^2 \cr}\eqno(25)$$
Fig.7 displays the threshold behaviour of $H \to W^+\overline{b}t$. The
branching ratio can be as large as $10^{-3}$ for $M_H < 2m_t$. If the
decay channel $H \to t\overline{t}$ is kinematically open one of course
has
$$\Gamma(H \to W^-\overline{b}t) \simeq \Gamma(H \to t\overline{t})
\eqno(26)$$
to a very high accuracy.
\bigskip
Before closing the section on Higgs decays let us give the branching ratio
of all three body decay modes considered above. We define
$$Br_{{}_{\{3body\}}}(M_H)=\displaystyle{\sum_{i=3body}\Gamma(H \to i)
\over \Gamma_{tot}^{Higgs}}\eqno(27)$$
Putting $m_t=150\ GeV$ and $E^{\gamma,g}_{cut}=20\ GeV$ we obtain
$$Br_{{}_{\{3body\}}}(M_H=400,\ 600,\ 800,\ 1000\ GeV)=(0.9,\ 3.5,\ 7.1,\ 9.4)
\cdot
10^{-2}\eqno(28)$$
Thus we see that the three body decay modes of a heavy Higgs can have
branching ratios as high as $10\%$.
\smallskip
Some comments on the feasibility to observe the discussed three body
Higgs decays are in order. We note that, out of the few
three body channels, $H \to W^+W^-(Z^0, \gamma)$ are the best
candidates to be seen at SSC energies ($\sqrt{s}=40TeV$) provided
$M_H \ge 600\ GeV$. The reason is that the background from the `direct'
production is [15] (for SSC)
$$\eqalign{&\sigma(pp \to W^+W^-Z^0+X) \simeq 0.4\ pb \cr
&\sigma(pp \to W^+W^- \gamma +X)|_{p^{\gamma}_T \ge 40\ GeV} \simeq 0.2
\ pb \cr}
\eqno(29)$$
whereas in the mass range $M_H=(600-1000)\ GeV$ one has [16]
$$\sigma(pp \to H+X)\simeq (10-1)\ pb \eqno(30)$$
Together with $Br(H \to W^+W^- (Z^0,\gamma))$ from figs.3 and 4 we find that
the three gauge bosons production through Higgs decay is `only' $10$
times smaller than the `direct' one ($q\overline{q'} \to W^+W^-Z^0$).
Therefore suitably chosen cuts should, in principle, make it possible
to observe such decays.
\smallskip
The situation for $H \to t\overline{t}(g,\gamma, Z^0)$ is a little
more involved. It is well known that there is an  overwhelming background
from $gg$ fusion [17]
$$\sigma(pp \to t\overline{t} +X) \simeq (10^4 -10^5)\ pb \eqno(31)$$
which makes it quite hard to observe hadronic Higgs decays in general.
The situation might be more promising for $V^0=\gamma, Z^0$ once
some suitable cuts are applied, but this would require more detailed
Monte Carlo simulations.
\smallskip
Let us also investigate the potential of the ambitious project of a
linear $e^+e^-$ super-collider operating at $\sqrt{s}=2TeV$ [18].
Here the heavy Higgs production is dominated by $W^*W^*$ and $Z^*Z^*$
fusion. Again for the range $M_H=(600-1000)\ GeV$ one has [18]
$$\eqalign{&\sigma (e^+e^- \to W^*W^* \to \nu \overline{\nu}H)
\simeq {\cal O}(10^{-1})\ pb \cr
&\sigma(e^+e^- \to Z^*Z^* \to e^+e^-H) \simeq {\cal O}(10^{-2}-10^{-3})
\ pb \cr}\eqno(32)$$
whereas the top production yields ($m_t=(100-200)\ GeV$) [19]
$$\eqalign{&\sigma(e^+e^- \to W^*W^* \to \nu \overline{\nu} t
\overline{t}) \simeq {\cal O}(10^{-2}-10^{-3})\ pb \cr
&\sigma(e^+e^- \to \gamma^* \gamma^* \to e^+e^- t \overline{t})
\simeq {\cal O}(10^{-1}-10^{-2})\ pb \cr
&\sigma_{beamstr.}(e^+e^- \to e^+e^- t\overline{t}) \simeq {\cal O}
(10-1)\ pb \cr}\eqno(33)$$
where the last cross section refers to beamstrahlung of electron
in the field of $e^+$-bunches (and vice versa) [20]. However,
designs of $e^+e^-$-machines which give rise to high large amount of
beamstrahlung seem to be disfavoured since such
designs would make the, usually `clean', $e^+e^-$-collider
`messier' due to possible underlying events caused by $\gamma \gamma$
interactions [21]. Keeping this
in mind it is clear that a high energy $e^+e^-$-collider has a better
potential to observe hadronic Higgs decays as compared to a $pp$-
machine. We also mention here that using more realistic
beamstrahlung spectra the cross section $\sigma_{beamstr.}$ is
smaller [21] as compared to the value we have quoted in (33).
\smallskip
\vfill\eject
{\bf 3. Dominant three body top decays}
\medskip
The relevant diagrams for the decays $t \to W^+b(\gamma ,g,Z^0)$ as well as
$t \to W^+bH$ are given in fig.8. As in the case of three body Higgs
decays (eq.(6)) the generic form of the partial width for the top decays
under consideration is
$$\Gamma(t \to 3body)=\displaystyle{1 \over 256 \pi^3 m_t^3}
\int_{s_2^-}^{s_2^+}ds_2 \int_{s_1^-}^{s_1^+}ds_1\overline{|T(t \to 3body)
|^2}\eqno(34)$$
Below we will give the new result on $t \to W^+bZ^0$.
Our calculations of
$t \to W^+b(\gamma ,g)$ confirms the results of ref.[9] and
the revised ones of ref.[8]. For completeness we
quote the numerical values for $\Gamma(t \to W^+bH)$ which has been
discussed in [7].

\bigskip
{\bf 3.1 $t \to W^+b(g, \gamma ,Z^0)$}
\medskip
The general form of the three amplitudes contributing to $t \to W^+bV^0$
($V^0=\gamma ,g,Z^0$) can be cast into the following expressions
$$\eqalign{&T_A^{WbV}=-i\displaystyle{gQ_t^V \over 2\sqrt{2}}
\displaystyle{\epsilon^{\nu *}(k_V) \epsilon^{\mu *}(k_W) \over s_2 -m_t^2}
\overline{u}_b \gamma_{\mu}\gamma_-\left(\lslash + m_t \right)\gamma_{\nu}
\left[g_V^t + g_A^t \gamma_5 \right]T^au_t \cr
&T_B^{WbV}=-i\displaystyle{gQ_b^V \over 2\sqrt{2}}
\displaystyle{\epsilon^{\nu *}(k_V)\epsilon^{\mu *}(k_W) \over s_1 - m_b^2}
\overline{u}_b\gamma_{\nu}\left[g_V^b + g_A^b \gamma_5 \right]T^a \left(
\rslash + m_b \right)\gamma_{\mu} \gamma_- u_t \cr
&T_C^{WbV}=-i\displaystyle{gg_{{}_{VWW}} \over 2\sqrt{2}}\epsilon^{\nu *}(k_V)
\epsilon^{\mu *}(k_W)
\displaystyle{\left
\{-g^{\lambda \rho} +q^{\lambda}q^{\rho}/M_W^2 \right\} \over s_3 -
M_W^2}f_{\rho
\mu \nu}\overline{u}_b \gamma_{\lambda} \gamma_- u_t \cr}\eqno(35)$$
where
$$\eqalign{\gamma_{\pm}&=1 \pm \gamma_5 \cr
f_{\rho \mu \nu}&=(k_W-k_V)_{\rho}g_{\mu \nu}-(2k_W+k_V)_{\nu}g_{\mu \rho}
+(2k_V+k_W)_{\mu}g_{\rho \nu}\cr}\eqno(36)$$
The coupling $g_{{}_{VWW}}$ has been defined in eq.(3) ($g_{{}_{gWW}}=0$) and
$Q_q^V$
are overall coupling constants of $V^0$ to quarks (charges in case of the
photon). Obviously for $V^0=\gamma ,g$ one has
$$g_V^q=1,\ \ \ g_A^q=0,\ \ \ q=t,b\eqno(37)$$
whereas for $V^0=Z^0$ we employ the common definitions
$$\eqalign{&g_V^t=\displaystyle{1 \over 4\cos \theta_W}\left[1 -
\displaystyle{8 \over 3}\sin^2 \theta_W \right], \ \ \ g_A^t=-
\displaystyle{1 \over 4\cos \theta_W} \cr
&g_V^b=\displaystyle{1 \over 4\cos \theta_W}\left[-1 + \displaystyle{4
\over 3}\sin^2 \theta_W \right], \ \ \ g_A^b=\displaystyle{1 \over 4
\cos \theta_W} \cr}\eqno(38)$$
The kinematical variables entering (35) are defined as follows
$$\eqalign{&l=k_W+k_b,\ \ \ l^2=s_2 \cr
&r=k_V+k_b,\ \ \ r^2=s_1 \cr
&q=k_W+k_V,\ \ \ q^2=s_3=M_W^2+M_V^2+m_t^2+m_b^2-s_1-s_2\cr}\eqno(39)$$
in which the phase space boundaries are readily found to be
$$\eqalign{s_2^-&=(M_W+m_b)^2,\ \ \ s_2^+=(m_t-M_V)^2 \cr
s_1^{\pm}&=M_V^2+M_W^2-\displaystyle{1 \over 2s_2}\bigg[\left(s_2-m_t^2
+M_V^2 \right)\left(s_2+M_W^2-m_b^2 \right) \cr
&\mp \lambda^{1/2}(s_2,M_W^2,m_b^2)\lambda^{1/2}(s_2,m_t^2,M_V^2)\bigg]
\cr}\eqno(40)$$
In case of $M_V=0$ we introduce a cut in the phase space on the
photon (gluon) energy
$$E^{\gamma ,g}=\displaystyle{m_t^2-s_2 \over 2 m_t} > E^{\gamma ,g}_{cut}
\eqno(41)$$
Since we keep $m_b \not = 0$ there are no collinear singularities.
\smallskip
We will not spell out here the expressions for the squared matrix
elements since for $V^0=\gamma ,g$ they can be found in [9] and the
corresponding formula for $V^0=Z^0$ is too lengthy. Instead we will
briefly comment on the factorization of the amplitude $T(t \to W^+b\gamma$
).
Splitting this amplitude into an abelian part which does not contain
gauge bosons couplings and a remainder $\tilde{T}$
$$\eqalign{&T(t \to W^+b\gamma)=-{1 \over 3}T_{abelian}+\tilde{T}\cr
&T_{abelian}=-i\displaystyle{ge \over 2\sqrt{2}}\epsilon_{\gamma}^{\nu *}
\epsilon_W^{\mu *}\overline{u}_b\bigg[\gamma_{\mu}\gamma_-
\displaystyle{(\lslash + m_t) \over -2k_{\gamma} \cdot k_t}\gamma_{\nu}+
\gamma_{\nu}\displaystyle{(\rslash +m_b) \over 2k_{\gamma} \cdot k_b}
\gamma_{\mu}\gamma_- \bigg]u_t \cr}\eqno(42)$$
one can show after some algebraic manipulation that the following
factorization holds
$$\tilde{T}=\left(\displaystyle{-2k_{\gamma} \cdot k_b \over 2k_{\gamma}
\cdot k_w}\right)T_{abelian} \eqno(43)$$
This fact simplifies the calculation to a large extent.
\smallskip
The numerical results for $Br(t \to W^+b(Z^0,\gamma,g))$ are plotted
versus the top mass in figs.9-11. Our results for photon and gluon
channel
confirm the ones
obtained in [9] and [23]. The decay $t\to Wb(\gamma ,g)$ as part of
radiative corrections to semi-leptonic top decays is also discussed in
[24]. The order of magnitude of these decays can be
summarized
by ($E^{\gamma, g}_{cut} < 20\ GeV$, $m_t=150\ GeV$)
$$\eqalign{&Br(t \to W^+bg) \simeq {\cal O}(10^{-1}) \cr
&Br(t \to W^+b \gamma) \simeq {\cal O}(10^{-3}) \cr}\eqno(44)$$
In SM with $m_t=200\ GeV$ we get
$$Br(t \to W^+bZ^0) \simeq {\cal O}(10^{-5}) \eqno(45)$$
where the suppression is essentially due to phase space.
There are, however, indeed very minor extensions of SM which allow values up to
$m_t=300\ GeV$ [22] (here the branching ratio reaches the $10^{-3}$ mark
).
\bigskip
{\bf 3.2 $t \to W^+ b H$}
\medskip
The decay $t \to W^+bH$ has been discussed by us in [7]. Here we
will just write down the relevant formulas.
\smallskip
Neglecting, as in the case $H \to W^-t\overline{b}$ (see eq.(24)), a third
possible diagram proportional to $m_b$ we are left with the two amplitudes
(see fig.8)
$$\eqalign{&T_A^{WbH}=-i\epsilon^{\mu *}(k_W)\displaystyle{g^2 \over
4\sqrt{2}}\
\displaystyle{m_t \over M_W}\displaystyle{V_{tb}^* \over s_1 -m_t^2}
\overline{u}_b \gamma_{\mu}\gamma_- \left(\lslash+m_t \right)u_t \cr
&T_B^{WbH}=-i\epsilon^{\mu *}(k_W)\displaystyle{g^2 \over 2\sqrt{2}}
\displaystyle{M_WV_{tb}^* \over s_2 -M_W^2}\overline{u}_b\gamma_+
\left[\gamma_{
\mu}-\displaystyle{m_t q_{\mu} \over M_W^2}\right] u_t
\cr}\eqno(46)$$
with
$$\eqalign{&l^2=(k_W+k_b)^2=s_1 \cr
&q^2=(k_H+k_W)^2=s_2 \cr}\eqno(47)$$
The reader will find the squared matrix element in the appendix
(eqs.(A4)-(A6)).
The
latter we write in the form
$$\overline{|T(t \to W^+bH)|^2}=\alpha_w^2 \pi^2|V_{tb}|^2\ \sum_{pol.}
 |\tilde{T}_A^{WbH} +
\tilde{T}_B^{WbH}|^2 \eqno(48)$$
The numerical results for this decay channel are displayed in fig.12.
Since the LEP data already restrict the Higgs mass $M_H \ge 50\ GeV$
[25] the branching ratio is expected to be small (like in the case $t
\to W^+bZ^0$). Indeed for $M_H=60\ GeV$ and $m_t=200\ GeV$ we obtain
$$Br(t \to W^+bH) \simeq {\cal O}(10^{-4}) \eqno(49)$$
We conclude by mentioning that the large number of $10^8$ $t\overline{t}$
pairs produced at SSC per year (which in case of hadronic Higgs decays gave
rise to an enormous background) is now of course of great advantage to
observe rare top decays (two body rare top decays like $t \to c(g,
\gamma ,H)$ have been investigated in [26]). Hence such decays would
be observable. In particular, if b-quark identification will be accessible
at high energies with high efficiency, then such three-body decays may be
probed by tagging three b-jet events in the five well-isolated jet
signals of the top decay.
\smallskip
We also note that with minor modifications (e.g. mixing angles) three
body decays of heavy fermions could become important in some extensions
of SM like two Higgs doublet models ($t \to W^+bH^0$) or fourth
generations extensions ($t' \to W^+bZ^0$). The latter is still not
excluded provided $m_{\nu_4} > M_Z/2$ (some models even favour such
an extension to explain the $\tau$-decay puzzle). The three body
decays with $Z^0$ bremsstrahlung of new particles have been discussed
in [27].
\bigskip\bigskip\bigskip
{\bf 4. Conclusions}
\medskip
The future high energy colliders like SSC or even a $2TeV$ $e^+e^-$-
machine will produce enough Higgs particles to look also into
subdominant decays of this yet missing ingredient of the Standard
Model. Here we have concentrated on the most important three body
decays of a heavy Higgs. We have shown that the branching ratio
$Br(H \to 3body)$ can be as high as $10\%$ and cannot therefore
be neglected, should indeed the Higgs turn out to be heavy. For the
channel $H \to W^+W^-Z^0$ we disagree with previous calculations.
This decay mode contributes substantially to the subclass of three
body decays of $H$.
\smallskip
In addition we have presented the dominant three body decays of
a heavy top quark. Some of these decays, like $t \to W^+bZ^0$, have
small (phase space suppressed) branching ratios. However, the
enormous number of $t\overline{t}$ pairs expected at SSC should make
it possible to observe even such rare decays provided the top quark
mass is around $200\ GeV$. Others, like $t \to W^+b(g,\gamma)$, with a
hard photon (gluon) have non-negligible branching ratios.

\bigskip\bigskip\bigskip
{\bf Acknowledgements.} We thank R.~Godbole
for valuable discussions and
comments. The
work of A.P. has been supported by a grant from the Postdoctoral
Graduate College of Germany and the work of M.N.
by Bundesministerium f\"ur Forschung und Technologie under the Grant Nr.
06KA757.
\vfill\eject

\centerline{\bf REFERENCES}\bigskip\bigskip
[1] For a summary see F.~Dydak in {\it Proceedings of the 25th
International Conference on High Energy Physics}, eds. K.~K.~Phua,
Y.~Yamaguchi \smallskip
[2] For a review see J.~F.~Gunion, H.~E.~Haber, G.~Kane and S.~Dawson,
{\it The Higgs Hunter's Guide} Addison Wesley 1990 \smallskip
[3] B.~W.~Lee, C.~Quigg and G.~B.~Thacker, Phys.~Rev.~{\bf D16} (1977)
1519 and references therein \smallskip
[4] T.~G.~Rizzo, Phys.~Lett.~{\bf B217} (1989) 191 \smallskip
[5] T.~G.~Rizzo, Phys.~Rev.~{\bf D31} (1985) 2366 \smallskip
[6] D.~A.~Dicus et al., Phys.~Rev.~{\bf D34} (1986) 2157 \smallskip
[7] R.~Decker, M.~Nowakowski and A.~Pilaftsis, Mod.~Phys.~Lett.~{\bf
A6} (1991) 3491; E {\bf A7} (1992) 819  \smallskip
[8] G.~Couture, Phys.~Rev.~{\bf D40} (1989) 2927;{\bf E42} (1990) 1855.
 \smallskip
[9] G.~Tupper et al., Phys.~Rev.~{\bf D43} (1991) 274 \smallskip
[10] K.~Hagiwara, K. Hikasa, R.~D.~Peccei and D.~Zeppenfeld, Nucl.~Phys.~{\bf
B282} (1987) 253 \smallskip
[11] P.~Langacker and J.~Liu, Philadelphia preprint UPR-0497T
(1992)\smallskip
[12] B.~Kniehl,  Nucl.~Phys.~{\bf B357} (1991) 439 \smallskip
[13] B.~Kniehl,  Nucl.~Phys.~{\bf B376} (1992) 3 \smallskip
[14] M.~Drees and K.~Hikasa,  Phys.~Lett.~{\bf B240} (1990) 455\smallskip
[15] V.~Barger and T.~Han, Phys.~Lett.~{\bf B212} (1988) 117
\smallskip
[16] H.~M.~Georgi et al., Phys.~Rev.~Lett.~{\bf 40} (1978) 692
\smallskip
G.~Altarelli, B.~Mele and F.~Pittolli, Nucl.~Phys.~{\bf B287} (1987)
205 \smallskip
[17] D.~Denegri in {\it Large Hadron Collider Workshop}, CERN 90-10,
Vol.1, 1990 \smallskip
[18] G.~Altarelli in {\it Proceedings of the Workshop on Physics
of Future Accelerators} CERN 87-07, Vol.1, 1987 \smallskip
[19] R.~P.~Kauffman, Phys.~Rev.~{\bf D41} (1990) 3343 \smallskip
[20] R.~Blankenbecler and S.~D.~Drell, Phys.~Rev.~{\bf D36} (1987)
277 \smallskip
[21] M.~Drees and R.~M.~Godbole, Phys.~Rev.~Lett.~{\bf 67} (1991)
1189; M.~Drees, R.~Godbole DESY-preprint 92-044
\smallskip
[22] S.~Bertolini, A.~Sirlin, Phys.~Lett.~{\bf B237} (1991) 179.
\smallskip
[23] V.~Barger, A.~Stange and W.-Y.~Keung, {\bf Phys.~Rev.~D42} (1990)
1835\smallskip
[24] M.~Jezabek and J.~H.~K\"uhn, Phys.~Lett.~{\bf B207} (1988) 91\smallskip
[25] D.~Decamp et al., ALEPH collab., Phys.~Lett.~{\bf B236} (1990) 233;
B.~Adeva et al., L3 collab., Phys.~Lett.~{\bf B248} (1990) 203;
P.~Aarnis et al., DELPHI collab., Phys.~Lett.~{\bf B245} (1990) 276;
M.~Z.~Akrawy et al., OPAL collab., Phys.~Lett.~{\bf B236} (1990) 224;
L3 collab., CERN-preprint CERN-PPE/92-40\smallskip
[26] G.~Eilam, J.~L.~Hewett and A.~Soni, Phys.~Rev.~{\bf D44} (1991)
1473 \smallskip
[27] V.~Barger, W.-Y.~Keung and T.~G.~Rizzo, Phys.~Rev.~{\bf D40}
(1989) 2274
\vfill\eject
{\bf Figure captions}
\medskip
{\bf Fig.1} Feynman diagrams contributing to $H \to W^+W^-V^0$,
$V^0 \in \{Z^0, \gamma \}$. In case of $V^0=\gamma$ clearly the
amplitude $T_C=0$
\medskip
{\bf Fig.2} Feynman diagrams relevant for the decay $H \to t
\overline{q}V$, $q \in \{t,b \}$ and $V \in \{Z^0 ,\gamma,g ,W\}$.
In case $H \to t\overline{t}(\gamma$ or $g)$ diagram C does not contribute.
For $H \to W^-t\overline{b}$ diagram B is negligible.
\medskip
{\bf Fig.3} Branching ratio $Br(H \to W^+W^-Z^0)$ versus the Higgs
mass.
\medskip
{\bf Fig.4} Branching ratio $Br(H \to W^+W^- \gamma)$ with three
different energy cuts: $E^{\gamma}_{cut}=10\ GeV$ (full line),
$20\ GeV$ (dashed), $50\ GeV$ (dotted).
\medskip
{\bf Fig.5} Branching ratio $Br(H \to t\overline{t}Z^0)$ versus
the Higgs mass with three different values of $m_t$: $m_t=90\ GeV$
(full line), $150\ GeV$ (dashed), $200\ GeV$ (dotted).
\medskip
{\bf Fig.6a} Branching ratio $Br(H \to t\overline{t}\gamma)$ with
$E^{\gamma}_{cut}=20\ GeV$. The full line corresponds to $m_t=100\ GeV$,
the dashed one to $150\ GeV$ and the dotted one to $200\ GeV$.
\medskip
{\bf Fig.6b} The same as Fig.6a but with $E^{\gamma}_{cut}=50\ GeV$.
\medskip
{\bf Fig.7} Branching ratio $Br(H \to t\overline{b}W^-)$ versus the
Higgs mass with three different values of the top mass indicated in
the figure. The dashed line corresponds to $Br(H \to t\overline{t})$.
\medskip
{\bf Fig.8} Feynman diagrams relevant for $t \to W^+bB$, $B \in
\{\gamma ,g, Z^0,H \}$. In case $B=g$ the diagram C does not
contribute.
In case of $t \to W^+bH$ diagram B is negligible.
\medskip
{\bf Fig.9} Branching ratio $Br(t \to W^+bZ^0)$ versus the top mass.
\medskip
{\bf Fig.10} Branching ratio $Br(t \to W^+b \gamma)$ with three
different energy cuts:$E^{\gamma}_{cut}=10\ GeV$ (full line), $20\ GeV$
(dashed), $50\ GeV$ (dotted).
\medskip
{\bf Fig.11} Branching ratio $Br(t \to W^+b g)$ with the same values
of the energy cuts as in fig.10.
\medskip
{\bf Fig.12} Branching ratio $Br(t \to W^+bH)$ versus the top mass
for $M_H=50\ GeV$ (full line), $60\ GeV$ (dashed), $70\ GeV$ (dotted).
\vfill\eject

{\bf Appendix}
\medskip
Below we give the expressions for different squared matrix elements.
It is convenient to define the following mass ratios
$$\eqalign{&\xi_H=\displaystyle{M_H^2 \over M_W^2}\cr
&\xi_t=\displaystyle{m_t^2 \over M_W^2} \cr}\eqno(A1)$$
We start with the decay $H \to W^+W^- \gamma$ (eq.(12)).
$$\eqalign{&\overline{|T(H \to W^+W^- \gamma)|^2}=
\displaystyle{e^2g^2M_W^2 \over 4}\bigg\{4M_W^2-M_H^2(2\xi_H-
\xi_H^2-14)\cr
&-2(s_1+s_2)(9+\xi_H^2)
+\displaystyle{s_1^2+s_2^2 \over M_W^2} (\xi_H+4)+\displaystyle{
2s_1s_2 \over M_W^2}\left( 2+2\xi_H -{s_1+s_2 \over M_W^2} \right) \cr
&+\left(\displaystyle{M_W^2-s_1 \over s_2 -M_W^2}\right)\bigg[
6M_W^2 - M_H^2(3-2\xi_H)
+s_1\left( 5-\xi_H -{2s_1+\xi_Hs_2 \over M_W^2}\right) \cr
&+s_2(5-3\xi_H-\xi_H^2)+{s_2^2 \over M_W^4} (2M_H^2-s_1-s_2)\bigg]\cr
&+\left(\displaystyle{M_W^2-s_2 \over s_1-M_W^2}\right)\bigg[
6M_W^2 -M_H^2(3-2\xi_H)+s_2\left( 5-\xi_H-{2s_2+\xi_H s_1\over M_W^2}
\right)\cr
&+s_1(5-3\xi_H-\xi_H^2)+{s_1^2\over M_W^4}(2M_H^2-s_1-s_2)\bigg]\bigg\} \cr}
\eqno(A2)$$
The three contributions (see eq.(21)) to $H \to t\overline{t}\gamma$
are
$$\eqalign{&|\tilde{T}_A^{t\overline{t}\gamma}|^2=\displaystyle
{1 \over (s_1-m_t^2)^2}\bigg[m_t^2(3m_t^2-3M_H^2+6s_1+s_2)
+s_1(M_H^2-s_1-s_2)\bigg]\cr
&|\tilde{T}_B^{t\overline{t}\gamma}|^2=\displaystyle{1 \over
(s_3-m_t^2)^2}\bigg[m_t^2(11m_t^2+M_H^2-2s_1-3s_2)
+s_1(M_H^2-s_1-s_2)\bigg]\cr
&2Re\tilde{T}_A^{t\overline{t}\gamma}(\tilde{T}_B^{t\overline{t}
\gamma})^*=\displaystyle{2 \over (s_1-m_t^2)(s_3-m_t^2)}\bigg[
m_t^2(7m_t^2-5M_H^2+2s_1-s_2)\cr
&+s_1(M_H^2-s_1-s_2)+M_H^2s_2\bigg]\cr}
\eqno(A3)$$
It is clear that the expressions for the squared matrix elements
of $H \to t\overline{b}W^-$ and $t \to W^+bH$ are related to each other
by a relative minus sign, which is a result of crossing symmetry and an
additional check for us. With obvious notation we will therefore write

$$\eqalign{&\left(\matrix{|\tilde{T}_A^{Wtb}|^2 \cr
|\tilde{T}_A^{WbH}|^2\cr}\right)=\mp \displaystyle{\xi_t \over (s_1-m_t^2)^2}
\bigg[ m_t^2(m_t^2-2s_2+s_1+M_H^2-6M_W^2)+2M_H^2M_W^2 \cr
&+s_1(\xi_ts_2+4\xi_ts_1-\xi_tM_H^2-2M^2_H-2M_W^2+2s_1+2s_2-
s_1s_2/M_W^2 ) \bigg] \cr}\eqno(A4)$$

$$\eqalign{&\left(\matrix{|\tilde{T}_C^{Wtb}|^2 \cr
|\tilde{T}_B^{WbH}|^2\cr}\right)=\mp
\displaystyle{1 \over (s_2-M_W^2)^2}\bigg[ m_t^2(9m_t^2-2\xi_tm_t^2M_H^2-
4s_1-4M_H^2\cr
&+8M_W^2-5s_2)+M_H^2(\xi_t^2M_H^2+4M_W^2)+4s_1(s_1-M_H^2-M_W^2+
\xi_tM_H^2)\cr
&+2s_2(-\xi_t^2M_H^2-\xi_tm_t^2+\xi_ts_2-2\xi_ts_1+\xi_tM_H^2+2s_1-
4M_W^2+\xi_t\xi_Hs_2)\cr
&+\xi_ts_2(\xi_ts_2-\xi_HM_H^2-s_2^2/M_W^2) \bigg] \cr}\eqno(A5)$$

$$\eqalign{&\left(\matrix{2Re\tilde{T}_A^{Wtb}(\tilde{T}_C^{Wtb})^*
\cr 2Re\tilde{T}_A^{WbH}(\tilde{T}_B^{WbH})^* \cr}\right)=
\mp \displaystyle{2\xi_t \over
(s_1-m_t^2)(s_2-M_W^2)}\bigg[ m_t^2(4m_t^2-3M_H^2+5M_W^2\cr
&-3s_2-2s_1)-2M_W^2(2s_2-M_H^2+2M_W^2)+s_1s_2(-2\xi_t-\xi_H+1+s_2/M_W^2)\cr
&+2s_1(\xi_tM_H^2+2s_1-M_H^2)+2M_H^2s_2 \bigg] \cr}\eqno(A6)$$
where the $\tilde{T}$'s enter eq.(23) for the Higgs decay and
(48) for the top decay.

\bye